\documentclass[aps,prb,twocolumn,superscriptaddress,showpacs,floatfix]{revtex4}
\usepackage{amsfonts}
\usepackage{amssymb}
\usepackage{amsmath}
\usepackage{graphicx}

\begin{document}

\title{Effects of magnetism and doping on the electron-phonon coupling in
BaFe$_{2}$As$_{2}$}
\author{L. Boeri}
\affiliation{Max-Planck-Institut f\"{u}r Festk\"{o}rperforschung, Heisenbergstra$\mathrm{%
\beta}$e 1, D-70569 Stuttgart, Germany}
\author{M. Calandra}
\affiliation{CNRS and Institut de Min\'eralogie et de Physique des Milieux Condens\'es,
case 115, 4 place Jussieu, 75252 Paris Cedex 05, France}
\author{I. I. Mazin}
\affiliation{Naval Research Laboratory, 4555 Overlook Avenue SW, Washington, DC 20375, USA}
\author{O.V. Dolgov}
\affiliation{Max-Planck-Institut f\"{u}r Festk\"{o}rperforschung, Heisenbergstra$\mathrm{%
\beta}$e 1, D-70569 Stuttgart, Germany}
\author{F. Mauri}
\affiliation{CNRS and Institut de Min\'eralogie et de Physique des Milieux Condens\'es,
case 115, 4 place Jussieu, 75252 Paris Cedex 05, France}
\date{\today}

\begin{abstract}
We calculate the effect of local magnetic moments on the electron-phonon coupling
in BaFe$_{2}$As$_{2}$+$\delta$ using the density functional perturbation theory. We show
that the magnetism enhances the total electron-phonon coupling by $\sim 50\%$,
up to $\lambda \lesssim 0.35$, still not enough to explain the high critical
temperature, but strong enough to have a non-negligible effect on
superconductivity, for instance, by frustrating the coupling with spin
fluctuations and inducing order parameter nodes. The enhancement comes
mostly from a renormalization of the electron-phonon matrix elements. We also
investigate, in the rigid band approximation, the effect of doping, and find
that $\lambda$ versus doping does not mirror the behavior of the density of
states; while the latter decreases upon electron doping, the former does
not, and even increases slightly.
\end{abstract}
\maketitle

The simultaneous presence of high T$_{c}$ superconductivity and magnetism in
the phase diagram of the Fe-based superconductors (FBSC) suggests that
magnetism plays an important role in determining the superconducting
properties. Phonons have been excluded early on as possible mediators, on
the basis of first-principles non-magnetic calculations for the undoped
compound, ~\cite{LFAO:DFT:boeri,LFAO:DFT:mazin} but the experimental situation is still
far from settled in this regard. ~\cite%
{phonons:isotope,phonons:raman,LFA:ARPES:Kordyuk}

Density functional (DFT) calculations correctly reproduce several properties
of Fe pnictides, such as the magnetic pattern in the parent compounds and
the geometry of the Fermi surface. 
The interplay of magnetic and elastic properties is however puzzling:~%
\cite{DFT:mazin:problems} on one hand, experiments measure weak magnetic
moments ($m\sim 0.3-1.0$ $\mu _{B}$) in the SDW state and no long-range
magnetic order in the 
 superconducting samples, while spin-polarized LSDA
calculations predict large magnetic moments at all dopings ($m\sim 2.0$ $\mu
_{B}$). On the other hand, the same spin-polarized LSDA calculations predict
equilibrium structures and phonon densities of states that are much closer
to the experiment than those predicted by non-magnetic calculations. ~\cite%
{BF2A2:DFT:yildirim,BaF2A2:DFT:zbiri,DFT:reznik:IXS} A
possible way to reconcile these two apparently conflicting results is an
itinerant picture, in which the Fe atoms nevertheless have large local
magnetic moments that order below the Neel temperature in the undoped
compound, but survive locally at all dopings.~\cite%
{FBSC:unusual:mazin,DFT:johannes:micro}. But this point of view immediately
raises the question of whether the estimate of electron-phonon ($e-ph$) matrix
elements given by early non-magnetic (rather than paramagnetic, $i.$ $e.,$
with the local Fe moment entirely suppressed) DFT calculations is
representative of the actual compounds.~\cite{LFAO:DFT:Yndurain}

In this paper, we calculate from first principles the $e-ph$
coupling constant in antiferromagnetic (AFM) and non-magnetic (NM) BaFe%
$_{2}$As$_{2}$, using the linear response method.~\cite{DFPT} We confirm
that magnetism strongly affects the phonon frequencies, leading to a
renormalization of the modes that involve Fe-As vibrations,~~\cite%
{BF2A2:DFT:yildirim,BaF2A2:DFT:zbiri,DFT:reznik:IXS} and
also find a strong effect on the $e-ph$ matrix elements, leading to a $\sim
50\%$ increase with respect to the NM values. Using a rigid-band model, we
show that the $e-ph$ coupling constant 
$\lambda $ as a function of doping does not follow the density of
states. Finally, we  estimate the $e-ph$
coupling in the \emph{paramagnetic} (PM) state, by combining the \textit{%
nonmagnetic} band structure (eigenenergies and eigenfunctions) with the 
\textit{magnetic} phonon spectra and  self-consistent potentials. For the relevant
values of doping, we estimate an upper bound to the $e-ph$ coupling constant 
$\lambda =0.35$, \emph{i.e.} not high enough to explain superconductivity,
but not sufficiently weak to be neglected. For instance, $e-ph$ interaction
may be one of the factors responsible for experimentally observed gap
anisotropy.%

\begin{center}
\begin{table}[tbh]
\begin{tabular}{|c|c|c|c|c|c|}
\hline
Magnetic Order & m ($\mu_{B}$) & $N(0)$(ev$^{-1}$) & $\lambda_{\sigma\sigma}$
& $\omega_{ln}$(K) & $\lambda_{\sigma\sigma}/N(0)$ \\ \hline
$NM$ & 0.0 & 1.18 & 0.18 & 194 & 0.15 \\ 
$AFMc$ & 2.4 & 1.36 & 0.33 & 179 & 0.24 \\ 
$AFMs$ & 2.6 & 0.68 & 0.18 & 180 & 0.26 \\ 
$PM1$ & $*$ & 1.18 & 0.27 & 206 & 0.23 \\ 
$PM2$ & $*$ & 1.18 & 0.27 & 195 & 0.23 \\ 
$PM3$ & $*$ & 1.18 & 0.31 & 170 & 0.26 \\ \hline
\end{tabular}%
\caption{Calculated properties of BaFe$_{2}$As$_{2}$. $m$ is the integral
over the cell of the absolute value of the magnetization, $N(0)$ is the DOS
in states/(spin eV $Fe$ atom), $\protect\lambda_{\protect\sigma\protect\sigma%
}$ is the electron-phonon coupling and $\protect\omega_{ln}$ is the
logarithmic-averaged phonon frequency. In addition to the three fully
self-consistent calculations we report here three model calculations (see
text): PM1 utilizes the wave functions, one-electron energies and phonon
frequencies from the NM calculations, and the deformation potentials from
the AFMc calculations. PM2 and PM3 use one-electron energies and
wavefunctions from the NM calculations, and phonon frequencies and
deformation potentials from the AFMc and AFMs calculations, respectively.}
\label{table:summary}
\end{table}
\end{center}

In order to disentangle the structural and magnetic effects, we used the
high-temperature tetragonal structure \cite%
{BaF2A2:struct:rotter,BaF2A2:crystal} in all our calculations.
In order
to estimate the effect of the SDW ordering vector on the
$e-ph$ properties,
along with
the NM calculations, we have considered two different AFM patterns: the
checkerboard one (AFMc), in which the nearest neighbor Fe spins are
antiparallel, and the experimentally observed one (AFM stripe, AFMs), in which spins are aligned
(anti)ferromagnetically along the ($y$)$x$ edge of the square Fe planes. 
%
The stabilization energies with respect to the NM solutions are
90 meV and 130 meV/Fe atom, respectively.

\begin{figure}[phb]
\centerline{\includegraphics*[width=0.9 \columnwidth]{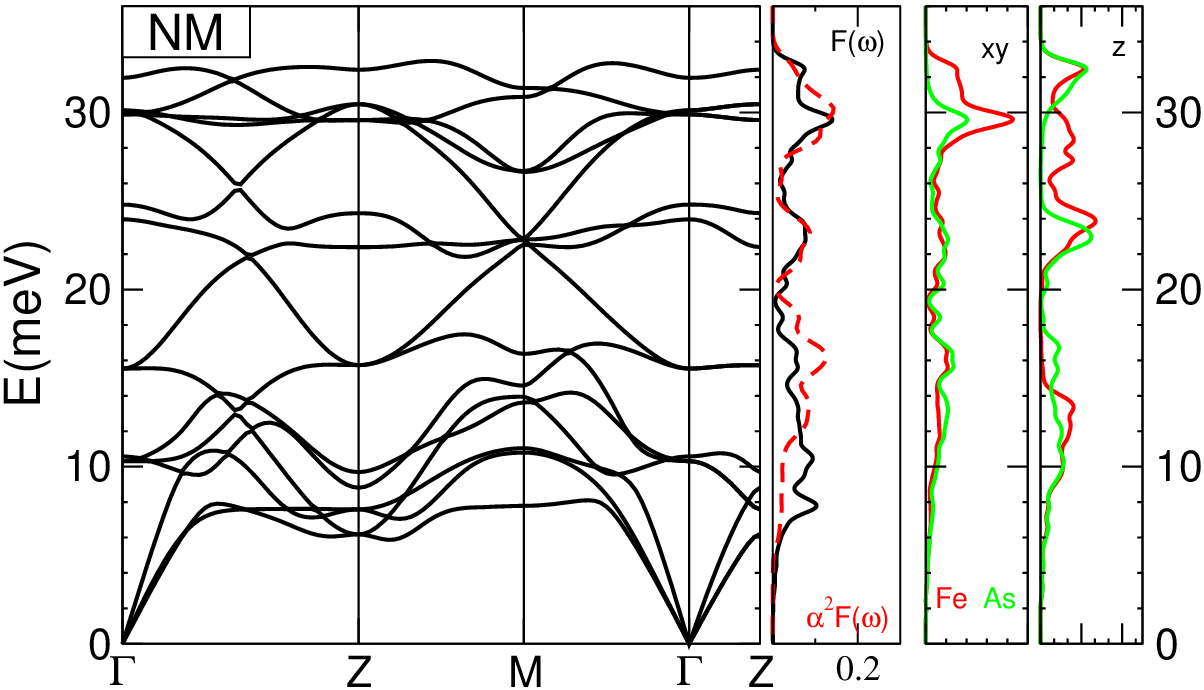}} 
\centerline{\includegraphics*[width=0.9 \columnwidth]{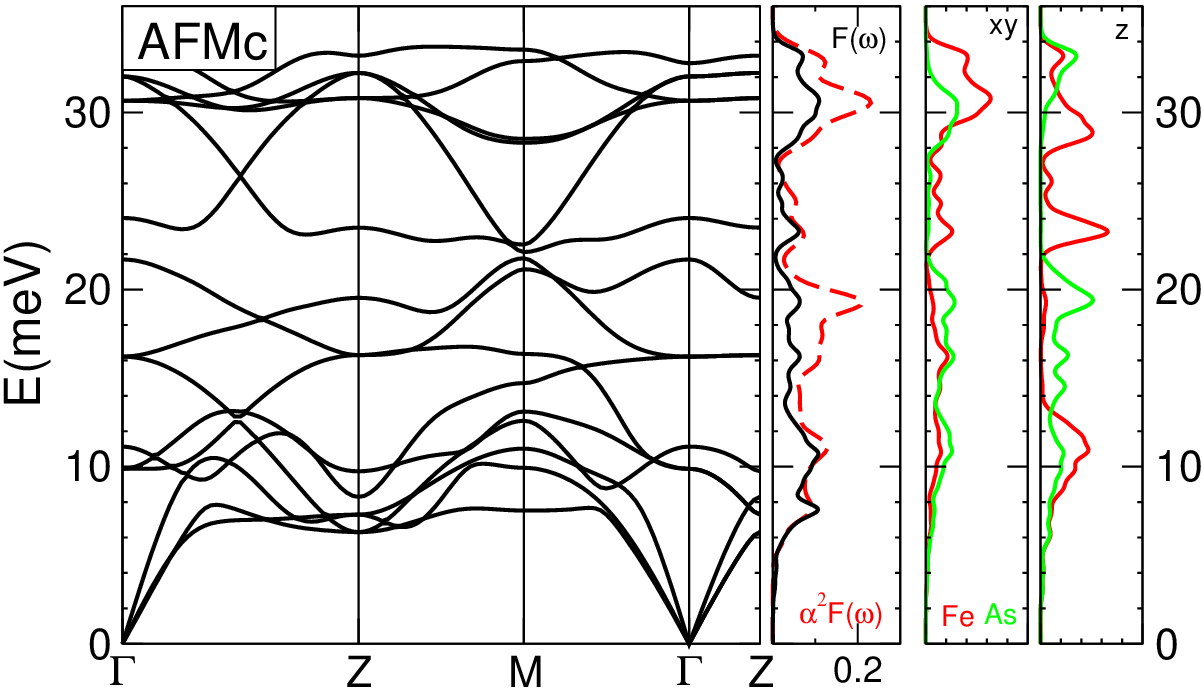}} 
\centerline{\includegraphics*[width=0.9 \columnwidth]{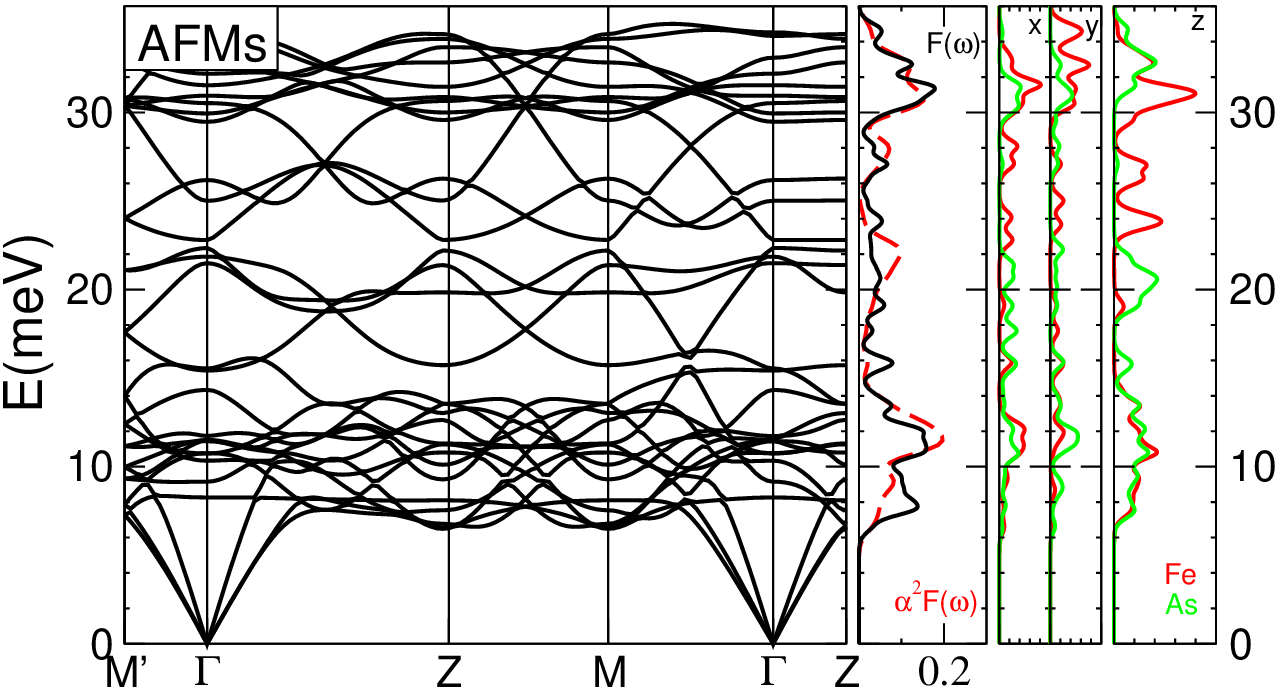}}
\caption{(color online)
Phonon properties of undoped BaFe$_2$As$_2$, for different
magnetic
patterns: non-magnetic (NM), AFM checkerboard (AFMc) and AFM stripe (AFMs).
From left to right: the phonon dispersions; the total phonon densities of states,
$F(\omega )$ (red, dashed), and Eliashberg function, $\alpha ^{2}F(\omega )$
(black, solid); partial contributions of Fe (red, dark) and As (green, light) ions
to the total $F(\omega )$, projected onto Cartesian axes (the Ba contribution is
limited to $\omega \leq 10$ meV, and not affected by magnetic order). The dispersions
are shown in the same Brillouin zone for the three patterns. The k-points are selected
so that they are physically the same in all three structures; the 
magnetic structure makes the two
$\Gamma-M$ directions inequivalent in the AFMs;
the spins are aligned AFM(FM) in the $x$($y$) direction.
  In this
coordinate system, the points are: $\Gamma =
(0, 0, 0)$;  $Z  = (0, 0, \pi/c)$ or $(\pi/a, \pi/a, 0)$; $ M  = (\pi/a, 0, 0)$;
$M' = (0, \pi/2a, 0)$, where $a$ is the length of the shortest Fe-Fe bond and $c$ is the
Fe-Fe interlayer distance.}
\label{fig:fig1}
\end{figure}
All calculations were performed in the generalized gradient approximation 
\cite{DFT:PBE} using plane-waves ~\cite{DFT:PWSCF} and ultra-soft
pseudopotentials~\cite{Vanderbilt}. We employed a cut-off of 40 (480) Ryd
for the wave-functions (charge densities). The electronic integration was
performed over an $8^{3}$ $\mathbf{k}$-mesh with a $0.01$ Ry
Hermitian-Gaussian smearing in the NM and AFMc case, while for the AFMs case
we used a $8\times 4\times 8$ $\mathbf{k}$-mesh with a $0.01$ Ryd
Hermitian-Gaussian smearing. Finer grids ($20^{3}$ and $20\times 10\times 20$
) were used for evaluating the densities of states (DOS) and the phonon
linewidths. Dynamical matrices and $e-ph$ linewidths were
calculated on $4^{3}$, NM and AFMc case, and $2^{3}$, AFMs case, uniform
grids in $\mathbf{q}$-space. Phonon frequencies throughout the Brillouin
Zone were obtained by Fourier interpolation. The (perturbed) potentials and
charge densities, as well as the phonon frequencies, were calculated
self-consistently at zero doping ($\delta =0$); the effect of doping on the $%
e-ph$ coupling was then estimated using the rigid-band
approximation.

The band structures at $\delta=0$ (not shown) agree with  previous
calculations.~\cite{BF2A2:DFT:singh,DFT:johannes:micro} The corresponding 
magnetic moments
and densities of states (DOS) at the Fermi level are reported
in table \ref{table:summary}.~\cite{foot1} The phonon dispersion for NM,
AFMc, and AFMs cases are shown in the left panels of Fig.~\ref{fig:fig1};
the remaining panels show the (partial) densities of states and Eliashberg
functions. Our results agree with previous calculations 
in the same crystallographic structures (adding orthorhombicity additionally
changes phonon dispersions, see Refs.~%
\onlinecite{BaF2A2:DFT:zbiri,DFT:reznik:IXS}). Magnetism has the biggest
effect on phonon modes that involve Fe-As vibrations; there is a pronounced
softening of a branch originally located at $\sim25$ meV along the $\Gamma-Z$
line in the NM calculation, down to $\sim20$ meV in both AFMc and AFMs
calculations. This branch corresponds to out-of-plane vibrations of the As
atoms. The in-plane Fe-As modes at low and high energy are sensitive not
only to the size, but also to the pattern, of the magnetic moment. Indeed,
when the order is AFMs, Fe vibrations along the AFM direction harden,
whereas those along the FM direction soften, while for the As vibrations it
is the opposite.

The shift of phonon frequencies in the AFM calculations has often been
considered an indication of an enhanced $e-ph$ coupling in the magnetic
phase,~\cite{LFAO:DFT:Yndurain,DFT:mazin:physC} but no explicit calculations
of the $e-ph$ coupling constant in the magnetic case have been reported so
far.
In this work, we have calculated from first principles the same-spin
component of the Eliashberg spectral function: 
\begin{eqnarray}
\alpha ^{2}F_{\sigma \sigma }(\omega ) &=&\frac{1}{N_{\sigma }(0)N_{k}}\sum_{%
\mathbf{k,q},\nu }| g_{\mathbf{k,k+q}}^{\nu ,\sigma }| ^{2}\times  
\notag \\
&&\delta (\varepsilon _{\mathbf{k}}^{\sigma })\delta (\varepsilon _{\mathbf{%
k+q}}^{\sigma })\delta (\omega -\omega _{\mathbf{q}}^{\nu }),
\label{eq:eliash:sc} \\
g_{\mathbf{k}n,\mathbf{k+q}m}^{\nu ,\sigma } &=&\langle \mathbf{k}^{\sigma}n|\delta
V_{\sigma }/\delta e_{\mathbf{q}\nu }|\mathbf{k+q}^{\sigma}m\rangle /\sqrt{2\omega _{%
\mathbf{q}\nu }}  \notag.
\end{eqnarray}%
Here, $N_{k}$ is the number of k-points used in the summation, $N_{\sigma
}(0)$ is the density of states per spin at the Fermi level, and $\omega _{%
\mathbf{q}}^{\nu }$ are the phonon frequencies. The $e-ph$ matrix
element $g_{\mathbf{k}n,\mathbf{k+q}m}^{\nu ,\sigma }$ is defined by the
variation of the self-consistent crystal potential $V_{\sigma }$  for the spin $\sigma $
with respect to a frozen phonon displacement according to the phonon
eigenvector $e_{\mathbf{q}\nu }=\sum_{A\alpha }M_{A}\sqrt{2\omega _{\mathbf{q%
}\nu }}\epsilon _{A\alpha }^{\mathbf{q}\nu }u_{\mathbf{q}A\alpha }$. Here $%
u_{\mathbf{q}A\alpha }$ is the Fourier transform of the $\alpha $ component
of the phonon displacement of the atom $A$ in the unit cell, $M_{A}$ is the
mass of atom $A$ and $\epsilon _{A\alpha }^{\mathbf{q}\nu }$ are $A\alpha $
components of $\mathbf{q}\nu $ phonon eigenvector normalized in the unit
cell. The first inverse moment of $\alpha ^{2}F(\omega )$ gives the
frequency-dependent $e-ph$ coupling constant: 
\begin{equation}
\lambda _{\sigma \sigma }(\omega )=2\int_{0}^{\omega }d\Omega \,\alpha
^{2}F_{\sigma \sigma }(\Omega )/\Omega .  \label{eq:lambda1}
\end{equation}%
The total $e-ph$ coupling constant $\lambda _{\sigma \sigma }=\lambda
_{\sigma \sigma }(\omega =\infty )$ is 0.18,0.33 and 0.18 for NM, AFMc and
AFMs, respectively. 

The three calculations in Fig.~\ref{fig:fig1} have different phonon spectra $%
\omega _{\mathbf{q}}^{\nu }$, different self-consistent crystal potentials $%
V_{\sigma }$, and different one-electron wave-functions $|\mathbf{k}n\rangle 
$ and eigenenergies $\varepsilon _{\mathbf{k}}^{\sigma }$, which all
determine the value of $\lambda $ in Eqs.~\ref{eq:eliash:sc}-\ref{eq:lambda1}%
. In order to understand which of these factors dominates the $e-ph$
coupling, we start for simplicity from the so-called Hopfield formula for
the $e-ph$ coupling constant: %
\begin{equation}
\lambda _{\sigma \sigma }=\frac{N_{\sigma }(0)D^{2}}{M\omega ^{2}},
\label{eq:Hopfield}
\end{equation}%
where $D$ is the deformation potential, which is a measure of the  average $e-ph$ interaction, and $M\omega ^{2}$ is a
characteristic force constant, which we assume depends weakly on the
magnetic order. We can then use the ratio $\lambda _{\sigma \sigma
}/N_{\sigma }(0)$ to obtain an estimate for the average change in the
$e-ph$ interaction due to magnetism; going from NM to AFMc (AFMs),
this ratio, and hence $D^{2}$, increases by $\sim 50\%$, from 0.15 to 0.24
(0.26) - see table \ref{table:summary}. %
A comparison of Eq.~\ref{eq:eliash:sc} and Eq.~\ref{eq:Hopfield} shows that
the increase in $D$ can be caused either by an increase in the derivative of
the DFT potential $V_{\sigma }$ with respect to displacement, or by a
difference in the one-electron wavefunctions, used to evaluate its
average over the Fermi surface. %
%
\begin{figure}[tbp]
\includegraphics*[width=0.95 \columnwidth]
{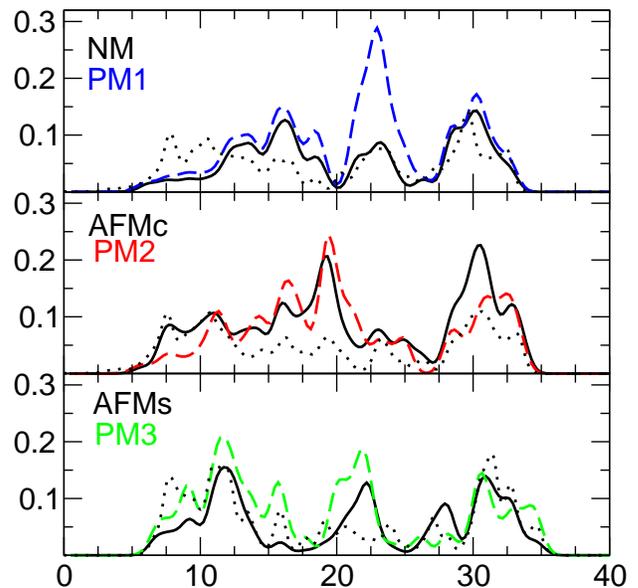}
\caption{({\em color online }:)
The Eliashberg functions
$\protect\alpha ^{2}F(\protect\omega )$
for the non-magnetic (NM, top, black solid line),
anti-ferromagnetic (AFM, middle and bottom, black solid lines) and model
paramagnetic calculations,
as described in the text, (PM1-3, colored dashed lines).
The corresponding phonon densities of states, in arbitrary units, are shown in each panel by the dotted lines.
}
\label{fig:fin}
\end{figure}
In order to disentangle these two effects, we performed a mixed calculation (PM1), in which we combined the NM eigenvalues and wavefunctions, as well as the
phonon frequencies, with the AFMc potential variations in the Eq.~\ref%
{eq:eliash:sc}. In the top panel of Fig.~\ref{fig:fin}, we compare the
resulting Eliashberg function $\alpha ^{2}F(\omega )$ with the NM one; the
total phonon DOS $F(\omega )$ is also shown as dotted lines. The effect of
using the AFM potential is an increased coupling of electrons to phonons
with frequencies $\omega \sim 20-25$ meV, while other modes are largely
unaffected. This increases $\lambda$  from 0.18 (NM) to 0.27 (PM1). The
middle and lower panel of Fig.~\ref{fig:fin} show two calculations, PM2 and
PM3, which use NM one-electron wave functions, and AFMc and AFMs phonon
frequencies and crystal potentials, respectively. A comparison of the $%
\alpha ^{2}F(\omega )$ with the phonon densities of states shows that, also
in these cases, there is an increased coupling to phonons around 20 meV,
both in AFM and PM. The $\lambda $'s, reported in table \ref{table:summary},
are 0.27 and 0.31, for PM2 and PM3 respectively. From the comparison of PM1
and PM2, we conclude that the effect of the phonon frequency is negligible,
while the $\sim 10\%$ spread of values between PM2 and PM3 gives an
indication of the effect of the short range AFM correlations on $\lambda $.
Finally, we find that the values of $\lambda _{\sigma \sigma }/N_{\sigma }(0)
$ in the PM calculations are in line with the AFM ones, indicating that, at $%
\delta =0$, the states at $E_{F}$ have a comparable, weak coupling to
phonons, for both NM and AFM orders. 

These model calculations, which combine the AFM potentials and phonons and
the NM wavefunctions and Fermi surfaces, represent the best approximation,
at LDA level, of the real \emph{para}magnetic (PM) state of
superconducting, doped, BaFe$_{2}$As$_{2}$ which is characterized by local,
disordered magnetic moments.~\cite{DMFT:Hansmann:moment} In the following,
we will use them to estimate an upper bound for the $e-ph$ coupling in the 
\emph{doped} BaFe$_{2}$As$_{2}$.

\begin{figure}[tbp]
\centerline{\includegraphics*[width=0.95 \columnwidth]{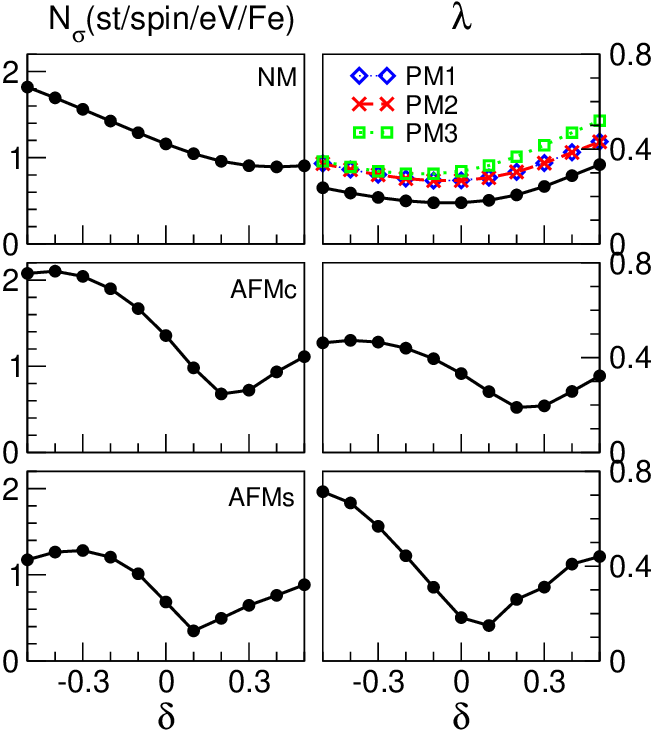}}
\caption{(color online) Effect of doping on the $e-ph$ coupling of BaFe$_2$As$_{2}$, in the rigid-band model.
 $\pm \delta$ is the number of excess electrons/holes per Fe atom.
 The left panel shows the density of states at $E_{F}$, in states/spin eV Fe atom. 
 The right panel shows the same-spin coupling constant
$\lambda =\lambda _{\sigma \sigma }$. $N_{%
\protect\sigma }(0)$ is the same for $NM$ and all three (see text) $PM$ calculations,
as described in text.}
\label{fig:dop}
\end{figure}

In Fig.~\ref{fig:dop}, we show the density of states af the Fermi level $%
N_{\sigma }(0)$ (\emph{left}) and $\lambda _{\sigma \sigma }$ (\emph{right})
as a function of doping $\delta $ for NM, AFMc, AFMs, and PM orders, from a
rigid-band calculation;
$\delta$ is defined as the number of excess electrons (holes) per Fe atom.
%
NM and PM calculations have the same $N_{\sigma }(0)$, which decreases
monotonically as a function of doping, while $\lambda _{\sigma \sigma }$ is
roughly symmetric around $\delta =0$. This indicates that, in Eqs.~\ref%
{eq:eliash:sc}-\ref{eq:Hopfield}, the effect of the matrix element dominates
over that of the DOS. In the AFM calculations, the shape of the $\lambda
_{\sigma \sigma }$ curve follows more closely that of $N_{\sigma }(0)$, with
a minimum at $\delta \sim 0.1-0.2$, and a maximum $\lambda _{\sigma \sigma
}=0.8$ for AFMs at $\delta =-0.5$,  corresponding to KFe$_{2}$As$_{2}$.
However, the use of the rigid-band approximation is questionable at 
these high dopings,
and in the following we limit our analysis to a smaller range
of $\delta = \pm 0.25$, where the error in $\lambda$ connected to the
rigid-band approximation is $\lesssim 20 \%$.~\cite{foot2}

We can summarize the results of Fig.~\ref{fig:dop}, by saying that, for 
$|\delta| < 0.25 $, the upper bound for the $e-ph$ coupling in 
paramagnetic, superconducting BaFe$%
_{2}$As$_{2}$ + $2 \delta$ is $\lambda _{\sigma \sigma }\lesssim 0.35$.
The enhancement in $\lambda $ results from a $\sim 50$ $\%$
increase in the $e-ph$ matrix elements, due to static magnetism, which is
independent on doping, and a symmetric increase of the matrix element for $%
\delta \neq 0$, which does not follow the shape of the electronic DOS.
%

The value  $\lambda _{\sigma \sigma }=0.35$,
 which includes magnetism and
doping, is almost a factor of two larger than that estimated in early
non-magnetic calculations for the undoped compounds. \cite%
{LFAO:DFT:boeri,LFAO:DFT:mazin}, and close to to a recent experimental estimate of $\lambda $
from a kink in photoemission spectra.~\cite{LFA:ARPES:Kordyuk}

 What are the consequences of this result?
Of course, it cannot explain a $T_{c}$ of 38 K,
which confirms that superconductivity in this and other Fe-based
superconductors is most likely due to electronic (magnetic) degrees of
freedom.~\cite{LFAO:DFT:mazin,LFAO:DFT:Kuroki}
However, in Fe pnictides, studies of the superconducting gap in realistic
models with AFM spin fluctuations show that solutions with and without gap
nodes are almost degenerate, so that even a relatively low $e-ph$ coupling
constant can help select either of them, by enhancing/suppressing the
pairing in the relevant channel.~\cite{DFT:mazin:physC,SF}

\textbf{Acknowledgements:} L.B. wishes to thank O.K. Andersen for
encouragement, support, and advice.  Part of the calculations were performed
at the IDRIS superconducting center.


\begin{thebibliography}{99}
\bibitem{BaF2A2:crystal}The lattice constants, as well as the
internal coordinate of the As atoms, are set to the experimental values:
$a=3.96 \AA$, $c=13.02 \AA$, $z_{As}=0.3545$.
\bibitem{foot1} Notice that the magnetic moment reported in table \ref%
{table:summary} is actually the integral of the absolute value of the
magnetization over the cell, and therefore larger than the moment inside the
Fe muffin-tin sphere, usually quoted in LAPW calculations.
\end{thebibliography}
\end{document}